\documentclass{INTERSPEECH2023}
\interspeechcameraready

\usepackage{graphicx}

\usepackage{colortbl}
\usepackage{color}
\usepackage{array}
\usepackage{amsmath, amssymb, scalerel}
\usepackage{nccmath}
\usepackage{booktabs}
\usepackage{tabularx}
\usepackage{algpseudocode}
\usepackage[table,dvipsnames]{xcolor}
\usepackage{todonotes}

\usepackage{arydshln}
\usepackage{url}
\usepackage{enumitem}

\usepackage{transparent}

\DeclareMathOperator*{\concat}{\scalerel*{\Vert}{\sum}}
\DeclareMathOperator*{\argmax}{arg\,max}

\usepackage{multirow}
\usepackage{xspace}
\makeatletter
\DeclareRobustCommand\onedot{\futurelet\@let@token\@onedot}
\def\@onedot{\ifx\@let@token.\else.\null\fi\xspace}
\def\eg{\emph{e.g}\onedot} 
\def\ie{\emph{i.e}\onedot}

\makeatother

\makeatletter
\def\adl@drawiv#1#2#3{%
        \hskip.5\tabcolsep
        \xleaders#3{#2.5\@tempdimb #1{1}#2.5\@tempdimb}%
                #2\z@ plus1fil minus1fil\relax
        \hskip.5\tabcolsep}
\newcommand{\cdashlinelr}[1]{%
  \noalign{\vskip\aboverulesep
           \global\let\@dashdrawstore\adl@draw
           \global\let\adl@draw\adl@drawiv}
  \cdashline{#1}
  \noalign{\global\let\adl@draw\@dashdrawstore
           \vskip\belowrulesep}}
\makeatother

\newcommand{\subtpm}[1]{\textsubscript{$\pm$#1}}

\newcommand{\um}[1]{\textcolor{black}{#1}}
\newcommand{\umb}[1]{\textcolor{black}{#1}}

\title{Online Continual Learning in Keyword Spotting for Low-Resource Devices\\via Pooling High-Order Temporal Statistics}

\name{Umberto Michieli, Pablo Peso Parada and Mete Ozay}
\address{Samsung Research UK}
\email{\{u.michieli, p.parada, m.ozay\}@samsung.com}

\begin{document}
\maketitle
\begin{abstract}

Keyword Spotting (KWS) models on embedded devices should adapt fast to new user-defined words without forgetting previous ones. Embedded devices have limited storage and computational resources, thus, they cannot save samples or update large models. We consider the setup of embedded online continual learning (EOCL), where KWS models with frozen backbone are trained to incrementally recognize new words from a non-repeated stream of samples, seen one at a time. To this end, we propose Temporal Aware Pooling (TAP) which constructs an enriched feature space computing high-order moments of speech features extracted by a pre-trained backbone. Our method, TAP-SLDA, updates a Gaussian model for each class on the enriched feature space to effectively use audio representations. In experimental analyses, TAP-SLDA outperforms competitors on several setups, backbones, and baselines, bringing a relative average gain of 11.3\% on the GSC dataset.

\end{abstract}
\section{Introduction}
\label{sec:intro}

\umb{Recently, deep networks have arrived on user devices to support popular tasks as Virtual Assistant (VA) interaction. In this paper, we focus on Keyword Spotting (KWS), which detects keywords in audio streams \cite{Rose1990,Chen2014} and plays a key role, \eg, to activate VAs with keywords \cite{lopez2021} or provide commands \cite{Tang2018}. Soon, users manifested the need for KWS models to recognize new user-defined words; however, despite the outstanding results of deep networks on standard benchmarks, they struggle to adapt.}

\umb{Continual Learning (CL)  emerged to adapt models to a target label space, unknown \textit{a-priori}. Its main challenge is {\it  forgetting} previous classes when learning new ones.
CL gained wide interest to avoid accessing personal data, however, most methods demand large computational  (\eg, computing Fisher matrix \cite{kirkpatrick2017overcoming}), or memory (\eg, storing replay data \cite{rebuffi2017icarl}) resources, which are unavailable on resource-constrained devices \cite{pellegrini2021continual,petit2023fetril}. 
Standard CL does not allow for fast adaptation. Ideally, a data point should be used immediately without storing it.
These problems are addressed by: (1) a single-epoch training, \ie, Online CL (OCL) \cite{mai2022}, to avoid storing samples of current task; (2) small batch sizes and (3) frozen backbone, \ie, OCL for Embedded systems (EOCL) \cite{hayes2022online,borghi2023challenges}, to comply with fast adaptation on low-resource devices. 
We focus on EOCL for KWS with no exemplar memory, where KWS models are partially updated with unitary batch size of data sampled from a non-repeated stream.}

\umb{
Non-online CL has been applied to KWS \cite{xiao2022}, combining data augmentation, replay and knowledge distillation; and ASR \cite{yang2022online}, where the term \textit{online} is used with a different meaning than in the current literature \cite{mai2022}, using 10 epochs.}
\cite{huang2022} presents a progressive CL method for KWS where different sub-networks are trained for different sets of keywords and a shared memory is used to store learnt features and perform knowledge transfer. Unlike our setup, \cite{huang2022} assumes task index known at inference time to select a proper sub-network. \cite{xiao2022} and \cite{huang2022} train all parameters with large batch sizes and pre-train the models on 50\% keywords. 
Standard CL methods (\eg, EWC \cite{kirkpatrick2017overcoming} and SI \cite{zenke2017continual}) are not applicable in our EOCL scenario because they assume: 1) training with large batch sizes; 2) storage of samples to perform many training epochs (not feasible on edge devices processing sensitive data); 3) updating whole feature extractors with many trainable parameters; 4) having one or more additional classification models for parameter regularization (\eg, through knowledge distillation). They perform poorly when evaluated in our setup \cite{mai2022}, hence they have been discarded.

\umb{Despite the progress on CL, EOCL for KWS remains unexplored in the literature, to the best of our knowledge.
OCL methods have been recently developed in vision tasks \cite{mai2022}, mainly based on memory \cite{rebuffi2017icarl,chaudhry2018efficient} and class prototypes \cite{mai2021supervised,hayes2022online}.}
\\
\umb{
In this work, we propose an effective approach for EOCL KWS without memory. Our contributions are summarized as follows: 
\begin{itemize}[noitemsep,topsep=0pt]
        \item We tackle the unsolved problem of EOCL for KWS for embedded devices with no storage and low computation power.
    \item Our method (TAP-SLDA) employs a new statistical pooling to extract enriched temporal information from speech features and a Gaussian-based classifier to model class representations on the enriched space with a shared covariance matrix, \ie, Streaming Linear Discriminant Analysis (SLDA) \cite{hayes2020lifelong}.
    \item TAP-SLDA outperforms multiple OCL methods and pooling mechanisms in a wide range of experiments, showing 9.9\% relative average gain compared to the best competitor. 
\end{itemize}
}

\begin{figure}[tb]
    \centering
    \includegraphics[trim=0cm 15.6cm 19.8cm 0cm, clip, width=\linewidth]{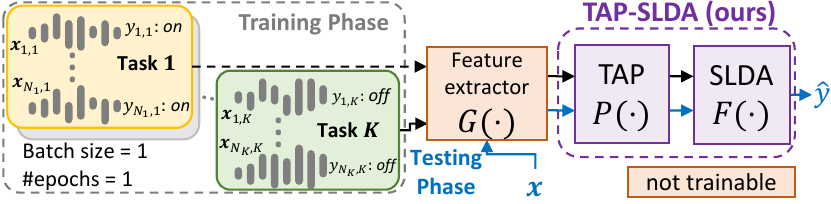}
    \caption{Our TAP-SLDA approach proposed for EOCL.}
    \vspace{-0.25cm}
\label{fig:TAPSLDA}
\end{figure}

\section{Our Method}
\label{sec:method}

The proposed method (Fig.~\ref{fig:TAPSLDA}) comprises: i) a feature extractor $G(\cdot)$, ii) a pooling mechanism $P(\cdot)$, and iii) a classifier $F(\cdot)$.
We incrementally train a model $F(P(G(\mathbf{x}_{n,k})))$ to generate predictions $\hat{y}_{n,k}$ via supervised OCL over $K$ subsequent tasks, where $\mathbf{x}_{n,k}$ is the $n$-th (${n=[N_k]}$) audio signal of the $k$-th task ($k=[K])$. 
$G(\cdot)$ is a pre-trained backbone (\eg, Wav2Vec2 \cite{w2v2}) and $F(\cdot)$ is a classifier.
Thus, we update $F(\cdot)$ in an embedded online fashion (\ie, processing one example at a time with no repetitions) using a fixed  $G(\cdot)$. \um{In Sect. 4, we consider the most challenging case of one single class per task.}

To perform pooling $P(\cdot)$, we propose a Temporal Aware Pooling (TAP) which computes and concatenates the first $R$ statistical moments from the output of $G(\cdot)$. Such moments characterize the distribution $\mathcal{G}$  of $g {\triangleq} G(\mathbf{x}_{n,k}) {\in} \mathbb{R}^{t\times d}$ where $t \times d$ is the temporal $\times$ feature size, and thus capture rich temporal statistics that improve keyword recognition. More formally,

\begin{equation}
    P\mkern-1.5mu\left(g\right) \mkern-2.5mu {=} \mkern-2mu  \concat \mkern-4.5mu  \left( \mkern-3mu  \mu, 
    {{E_{\mathcal{G}}\mkern-5mu\left[  (g {-}\mu)^2  \right]^\frac{1}{2}}}\mkern-1.5mu, \mkern-4.5mu
    \concat_{r=3}^R \mkern-5mu E_{\mathcal{G}}\mkern-5mu\left[\mkern-1.5mu
    \frac{g{-}\mu}
    {E_{\mathcal{G}}\left[  (g {-}\mu)^2  \right]^\frac{1}{2}}
    \mkern-1.5mu\right]^{\mkern-1.5mur}
    \right)\mkern-2.5mu,
\end{equation}
where $E_{\mathcal{G}}[\cdot]$ is the expectation on ${\mathcal{G}}$, $\mu$ the empirical mean, and $\concat$ the concatenation. That is, $P(g) {\in} \mathbb{R}^{R \cdot d}$ concatenates the first $R$ moments of $g$.
\um{In speaker verification: \cite{rouvier2021study} shows that 3$^\mathrm{rd}$-4$^\mathrm{th}$ moments alone are not useful; \cite{you2019multi} uses $R\!=\!4$ for auxiliary tasks.
In our case, we feed $R\!=\!5$ moments to the  classifier.}

\begin{figure}[tb]
    \centering
    \includegraphics[trim=0cm 17.9cm 27.8cm 0cm, clip, width=\linewidth]{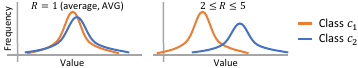}
    \caption{Distribution of statistical moments of $\mathcal{G}$ for $c_1 {\neq} c_2$.}
    \vspace{-0.2cm}
\label{fig:intuition}
\end{figure}

Lastly, we develop a classifier $F(\cdot)$ via SLDA \cite{hayes2020lifelong}. This classifier computes a running mean feature vector per class (\ie, prototypes) and one covariance matrix shared across classes, that is updated online via \cite{dasgupta2007line}. During inference, SLDA assigns the label of the  Gaussian model of a category to the input sample.
\umb{Our TAP-SLDA extracts an enriched feature space with different temporal statistics, which
encode information about evolution of phonetic sounds within each word \cite{w2v2}. Consequently, the  covariance matrix shared among classes allows to identify common relationships between the temporal statistics among different words which helps to recognize new unseen words.
We hypothesise that features of different classes have similar distribution of 1$^{\mathrm{st}}$ moments, while higher moments capture the difference, as illustrated in Fig.~\ref{fig:intuition} and quantified in Sec.~\ref{sec:results}.}

\textbf{OCL baselines} to update $F(\cdot)$ are described next.
\noindent\textbf{Fine-Tuning (FT)} updates a linear output layer using stochastic gradient descent and cross-entropy loss with no mechanisms to prevent forgetting.
\noindent\textbf{Online Perceptron (PRCP)} keeps one weight vector for each class \cite{hayes2022online}, initialized to the first sample of the class. After that, when the model misclassifies a new sample, the class weight vector and the weight vector of the misclassified class with the highest score are updated. During inference, the label is assigned by taking the $\argmax$ over the dot product between weights and input vector.
\noindent\textbf{Nearest Class Mean (NCM)} stores a class-wise running mean feature vector
\cite{rebuffi2017icarl,mai2021supervised,hayes2022online,mensink2013distance}. %
At inference on a new sample, NCM assigns the label of the nearest centroid using Euclidean distance.
\noindent\textbf{Online CBCL} extends NCM with multiple class centroids \cite{ayub2020cognitively}. At inference, CBCL searches for the weighted nearest neighbor, where class weights are inversely proportional to the number of samples seen so far for that class.
We use the default parameters suggested in \cite{ayub2021f}.
\noindent\textbf{Streaming One-vs-Rest (SOvR)} measures how close a new input is to a class mean vector while also considering its distance to examples from other classes \cite{hayes2022online}, which is reminiscent of SVMs. 
\noindent\textbf{SQDA} extends SLDA estimating one covariance matrix for each class (\ie, Quadratic). The drawbacks are increased memory consumption and need for many samples per class to estimate reliable covariance matrices \cite{anagnostopoulos2012online}.
\noindent\textbf{Streaming Gaussian Na\"ive Bayes (SNB)} estimates a running variance vector per class \cite{welford1962note} (\ie, diagonal covariance matrices assuming independent features). It requires significantly less memory than SQDA.
\noindent\textbf{Online iCaRL} stores a class-balanced memory buffer 
\cite{rebuffi2017icarl,castro2018end,hayes2020lifelong}
and randomly replaces an example from the most represented class with a new sample when the buffer is full. During training, it randomly draws examples from the buffer, combines them with the new sample and updates a single linear output layer. 
While effective, it stores sensitive samples.

\textbf{Pooling baselines} compared against our TAP are described next:
\noindent\textbf{AVG} is average pooling \cite{lecun1998gradient};
\noindent\textbf{MAX} is max pooling \cite{riesenhuber1999hierarchical};
\noindent\textbf{MIX} takes a linear combination of AVG and MAX \cite{zhou2021mixed};
\noindent\textbf{STOCHASTIC} selects activations based on probabilities and further calculates output based on multinomial distribution \cite{zeiler2013stochastic}, inspired by dropout; 
\noindent\textbf{RAP} concatenates top-k\% features to discover information spread over the time \cite{bera2020effect};
\noindent\textbf{AVGMAX} concatenates AVG and MAX pooling \cite{monteiro2020performance};
\noindent\textbf{TSTP} concatenates first and second order moments \cite{wang2021revisiting};
\noindent\textbf{TSDP} considers only the second moment \cite{wang2021revisiting};
\noindent\textbf{L}$_{\mathbf{p}}$\textbf{-norm} computes average of $p$-th order non-central moment \cite{feng2011geometric}, while we employ high order central moments;
\noindent and \textbf{iSQRT-COV} computes second-order covariance pooling via iterative  matrix square root normalization \cite{li2018towards}.

\section{Implementation Details}
\label{sec:implementation}
\noindent\textbf{Datasets.}
\um{We consider two datasets. Each task includes only keywords of the same class (\ie, samples are sorted by label).}
\textit{GSC v2} \cite{gsc} comprises %
$\sim$100k utterances and 35 different English words. The number of elements per word is unbalanced ranging from 1.5k to 4k. %
\textit{MSWC} \cite{mswc} contains 23.4M spoken words and over 340k different words in 50 languages. We picked the 5 most represented languages (\ie, \textit{en}, \textit{fr}, \textit{ca}, and \textit{rw}) and we created 3 micro-sets for each language with different number of words ${N=\{25,50,100\}}$.
Each micro-set includes the $N$ most frequent words, randomly selecting 1k samples for each word. \textit{Dev} and \textit{test} sets are also filtered for each case to include only the $N$ words present in the \textit{train} set \cite{splits}. 

\noindent Unless otherwise stated, we consider a class-IID setup, where all samples are ordered by class and shuffled within each class.

\noindent\textbf{Models.}
We use 6 recent speech recognition backbones as $G(\cdot)$: Wav2Vec2-B/L \cite{w2v2}, HuBERT-B/-L/-XL \cite{HuBERT} and Emformer-B \cite{emformer}.
All models are pre-trained \cite{torchaudio} on 
LibriSpeech \cite{panayotov2015librispeech}.

\begin{table*}[ht]
\setlength{\tabcolsep}{1.75pt}
  \centering
  \caption{Results of $10$ OCL methods and $6$ backbones on the GSC dataset. Each entry is averaged over $5$ distinct class ordering.}
  \resizebox{\linewidth}{!}{%
    \begin{tabular}{lcccccccccccccccccccccccccccc}
    \toprule
    \multicolumn{1}{r}{} & \multicolumn{4}{c}{W2V-B} & \multicolumn{4}{c}{W2V-L} & \multicolumn{4}{c}{Emf-B} & \multicolumn{4}{c}{HB-B} & \multicolumn{4}{c}{HB-L} & \multicolumn{4}{c}{HB-XL} \\
    \cmidrule(lr){2-5} \cmidrule(lr){6-9} \cmidrule(lr){10-13} \cmidrule(lr){14-17} \cmidrule(lr){18-21} \cmidrule(lr){22-25} 
    \multicolumn{1}{r}{} & Acc & BwT & Forg & Pla & Acc & BwT & Forg & Pla & Acc & BwT & Forg & Pla & Acc & BwT & Forg & Pla & Acc & BwT & Forg & Pla & Acc & BwT & Forg & Pla  \\\midrule
    
    FT    & 2.5\subtpm{1.0} & 1.2 & 98.5 & \textbf{100}  & 2.5\subtpm{1.0} & 1.3 & 98.5 & \textbf{100}  & 2.7\subtpm{0.8} & 1.5 & 97.6 & \textbf{99.8} & 2.5\subtpm{1.0} & 1.2 & 98.5 & \textbf{100}  & 2.5\subtpm{1.0} & 1.2 & 98.5 & \textbf{100}  & 2.5\subtpm{1.0} & 1.2 & 98.5 & \textbf{100}   \\
    
    PRCP \cite{hayes2022online}  & 2.5\subtpm{1.0}& 1.2 & 98.4 & \textbf{100} & 2.5\subtpm{1.0} & 1.3 & 98.4 & \textbf{100}   & 2.9\subtpm{0.7}  & 2.2 & 95.8 & 97.5     & 2.5\subtpm{1.0}  & 1.4 & 98.4 & \textbf{100}  & 2.5\subtpm{1.0} & 1.2 & 98.4 & \textbf{100}   & 2.6\subtpm{1.0}  & 1.3 & 98.4 &\textbf{100}  \\
    
    SNB \cite{welford1962note}  & 3.7\subtpm{0.0} & 6.5 & 38.6 & 15.7 & 9.4\subtpm{2.4} & 9.2 & 36.1 & 16.6  & 7.2\subtpm{0.0}  & 13.1 & 24.8 & 22.8 & 77.4\subtpm{0.0} & 80.9 & 18.1 & 84.2 & 6.5\subtpm{0.0}  & 12.0 & 37.1 & 23.7 & 57.9\subtpm{0.0}  & 60.3 & 23.8 & 65.2  \\
    SOvR \cite{hayes2022online}  & 1.8\subtpm{0.0} & 6.2 & 39.6 & 14.5   & 1.8\subtpm{0.0}  & 6.2 & 31.6 & 15.8   & 4.9\subtpm{0.0}   & 7.3 & 24.4 & 14.5  & 19.1\subtpm{0.0} & 29.8 & 41.2 & 42.5 & 14.0\subtpm{0.0} & 21.8 & 34.8 & 32.9     & 15.7\subtpm{0.0} & 22.6 & 35.3 & 33.8    \\
    
    NCM \cite{mensink2013distance}  & 67.5\subtpm{6.0}  & 74.0 & 13.2 & 77.3 & 69.5\subtpm{7.0}  & 76.0 & 12.1 & 80.0 & 8.8\subtpm{0.0}  & 14.1 & 23.6 & 22.0  & 83.9\subtpm{0.0} & 86.0 & 8.2 & 89.1   & 46.7\subtpm{0.0} & 55.4 & 22.8 & 62.4   & 62.5\subtpm{0.0}  & 66.7 & 17.3 & 72.2  \\
    
    SLDA \cite{hayes2020lifelong} & 82.4\subtpm{0.1} & 84.2 & 8.0 & 87.0 & 81.6\subtpm{0.1} & 83.8 & 8.3 & 86.8 & 23.2\subtpm{0.0}  & 32.9 & 23.7 & 41.9 & 94.2\subtpm{0.0} & 94.9 & 3.8 & 96.2 & 85.5\subtpm{0.0} & 88.2 & 8.3 & 90.8 & 93.3\subtpm{0.0} & 94.1 & 5.1 & 95.4  \\
    SQDA \cite{anagnostopoulos2012online} & 80.6\subtpm{2.5} & 78.2 & 5.7 & 81.2 & 80.5\subtpm{2.4} & 76.9 & \textbf{5.1} & 80.6 & 24.3\subtpm{0.7} & 21.9 & \textbf{17.3} & 31.5 & 90.0\subtpm{3.4} & 87.8 & \textbf{2.8} & 90.0 & 67.4\subtpm{4.4} & 59.4 & \textbf{4.4} & 64.8 & 83.0\subtpm{2.1} & 73.9 & \textbf{0.0} & 76.8 \\
    
    \multirow{2}{*}{\shortstack[l]{TAP-SLDA\\\footnotesize(ours)}}  & \multirow{2}{*}{\shortstack[l]{\textbf{89.9}}\subtpm{0.0}}& \multirow{2}{*}{\shortstack[l]{\textbf{91.8}}} & \multirow{2}{*}{\shortstack[l]{\textbf{5.6}}} & \multirow{2}{*}{\shortstack[l]{93.7}}  & \multirow{2}{*}{\shortstack[l]{\textbf{90.0}\subtpm{0.0}}}& \multirow{2}{*}{\shortstack[l]{\textbf{91.7}}} & \multirow{2}{*}{\shortstack[l]{5.4}} & \multirow{2}{*}{\shortstack[l]{93.4}}  & \multirow{2}{*}{\shortstack[l]{\textbf{50.8}\subtpm{0.3}}}& \multirow{2}{*}{\shortstack[l]{\textbf{58.8}}} & \multirow{2}{*}{\shortstack[l]{20.3}} & \multirow{2}{*}{\shortstack[l]{65.8}}  & \multirow{2}{*}{\shortstack[l]{\textbf{95.7}\subtpm{0.0}}} & \multirow{2}{*}{\shortstack[l]{\textbf{96.0}}} & \multirow{2}{*}{\shortstack[l]{3.0}} & \multirow{2}{*}{\shortstack[l]{96.9}} & \multirow{2}{*}{\shortstack[l]{\textbf{90.8}\subtpm{0.0}}} & \multirow{2}{*}{\shortstack[l]{\textbf{91.8}}} & \multirow{2}{*}{\shortstack[l]{6.1}} & \multirow{2}{*}{\shortstack[l]{93.6}} & \multirow{2}{*}{\shortstack[l]{\textbf{95.5}\subtpm{0.0}}} & \multirow{2}{*}{\shortstack[l]{\textbf{95.8}}} & \multirow{2}{*}{\shortstack[l]{3.4}} & \multirow{2}{*}{\shortstack[l]{96.6}}   \\
    
    & & \\\hdashline

    iCaRL \cite{rebuffi2017icarl}  & 76.9\subtpm{1.0} & 79.1 & 14.7 & 83.6  & 73.6\subtpm{1.8} & 78.0 & 17.4 & 83.5 & 18.2\subtpm{0.3} & 26.9 & 28.8 & 44.7  & 93.7\subtpm{0.1} & 94.6 & 4.2 & 96.7   & 78.5\subtpm{0.5} & 83.1 & 12.8 & 85.1   & 92.9\subtpm{0.3} & 93.8 & 5.3 & 95.5   \\
    \midrule
    \textbf{Avg} & 47.5 & 49.6 & 33.5 & 73.0 & 48.1 &50.1 & 32.5 & 73.7 & 15.1 & 19.3 & 38.0 & 46.2 & 64.3 & 65.9 & 28.6 & 88.5  & 44.1 & 46.9 & 34.6 & 71.6 & 56.8 & 57.7 & 30.4 & 80.8  \\\bottomrule
\end{tabular}%
}
  \label{tab:results_main_table_online_cl_methods_bis}%
   \vspace{-0.2cm}
\end{table*}%

\noindent\textbf{Metrics.}
We evaluate plasticity and forgetting of EOCL methods \cite{mai2022} on test sets, at the end of training.
We compute accuracy (Acc \%, $\uparrow$) to assess the final model performance on all classes. \um{\um{We compute the relative gain  of $\mathrm{Acc}_2$ compared to $\mathrm{Acc}_1$ with respect to the available room of improvement by ${(\mathrm{Acc}_2- \mathrm{Acc}_1) / (100 - \mathrm{Acc}_1)}$.}}
We measure: 
(1) stability by backward transfer (BwT, $\uparrow$), which tracks the influence that learning a new task has on the preceding tasks’ performance; 
(2) forgetting (Forg, $\downarrow$), which averages the difference of class-wise Acc achieved at the last step and the best class-wise Acc achieved previously; (3) plasticity (Pla, $\uparrow$) as the average Acc achieved on each task evaluated right after learning that task \cite{huang2022}.
Memory usage is discussed directly in Sec.~\ref{sec:results}.

\section{Experimental Results}
\label{sec:results}

All results are averaged over $5$ class orderings, although standard deviation is not always shown for readability.
Offline bounds are not shown since $F(\cdot)$ is different for each approach.

\noindent\textbf{TAP-SLDA vs.\ OCL methods} 
is shown in Tab.~\ref{tab:results_main_table_online_cl_methods_bis}. Some methods (FT and PRCP) largely overfit to the last seen data and cannot mitigate forgetting (low Acc/BwT and high Forg/Pla). SNB and SOvR improve on HB-B and XL thanks to the centroid-based inference mechanism (Pla dropped significantly).
NCM and CBCL achieve the same results  due to the default parameter configuration of CBCL suggested in \cite{ayub2020cognitively}. They improve results on all backbones (medium Acc/BwT/Forg and high Pla), except for the memory-efficient Emf-B.
SLDA significantly improves final performance thanks to the shared covariance matrix among classes (high Acc/BwT/Pla and low Forg). SQDA achieves lower results than SLDA: it can preserve previous tasks (lowest Forg, \ie, good stability), however, it cannot adapt well to the new tasks (Pla lower than SLDA, \ie, worse plasticity). Storing one covariance matrix per each class makes SQDA not only storage demanding, but also ineffective if data is scarce and reliable statistics cannot be extracted for each class, as highlighted by high standard deviation.
Hence, we selected SLDA as the best method, and we applied our TAP pooling on top of it (\ie, TAP-SLDA). TAP-SLDA shows strong Acc gains on every backbone, improving on average by relative 
\um{37.8\%} 
compared to SLDA (76.7\% vs.\ 85.5\%). On Emf-B, the relative gain is 
\um{35.0\%} 
compared to the runner-up SQDA, and 
\um{25.9\%} 
compared to the baseline SLDA. Looking at BwT, Pla and Forg metrics, we observe that our TAP-SLDA achieves the best trade-off between plasticity in learning new tasks and forgetting past ones. The key element in TAP-SLDA is the computation of the high-order feature space that provide useful temporal characteristics to the subsequent centroids- and covariance- based modeling. \um{That is, providing richer, temporal-aware statistics of  input waveform is beneficial for EOCL KWS, where models must adapt fast exploiting all the information and relate it to previous knowledge.}

Finally, we use a simple replay method, iCaRL,
 with 1k buffer size. iCARL is competitive, however: 1) it doubles the batch size as one new sample is coupled with a replay one, 2) it stores sensitive data, thus raising privacy/storage concerns, 3) our method surpasses it by large margin 
(85.5\% vs.\ 72.3\%).

\noindent\textbf{TAP improves every OCL method.} To show benefits of TAP to EOCL KWS, we include it on top of each OCL method in Tab.~\ref{tab:results_main_table_online_cl_methods}. Comparing the results against Tab.~\ref{tab:results_main_table_online_cl_methods_bis} 
emerges that TAP improves OCL methods almost every time (7 exceptions out of 54 cases). Sometime, we observe a large gain up to $\sim$60\%.
\umb{This is the consequence of using high-order statistics.
As preluded in Fig.~\ref{fig:intuition}, we compute the mean Wasserstein distance between distributions of features of pair-wise classes to quantify this effect. The distance of the $1^{\mathrm{st}}$ moments
($0.09 \pm 0.02$) is considerably lower than the mean distance of the 2$^\mathrm{nd}$ to the 5$^{\mathrm{th}}$ moments ($0.22 \pm 0.07$) 
indicating that classes are more entangled in the feature space of the 1$^{\mathrm{st}}$ moment (low distance) than of higher moments. Thus, high moments improve the classification score.}

\begin{table}[tbp]
\setlength{\tabcolsep}{1.7pt}
  \centering
  \caption{Acc of OCL methods employing our TAP on GSC. Improvement with respect to original methods within brackets.}
    \resizebox{0.95\linewidth}{!}{%
    \begin{tabular}{lccccccc}
    \toprule
    \textbf{TAP+} & W2V-B & W2V-L & Emf-B & HB-B & HB-L & HB-XL & \multicolumn{1}{c}{\textbf{Avg}} \\
    \midrule
    FT    & 5.4  & 6.1  & 2.9    & 2.7   & 2.7   & 2.7    & \multicolumn{1}{c}{3.8 \color{OliveGreen}\footnotesize(+1.2)}  \\
    PRCP & 3.5   & 4.6    & 3.0  & 2.8   & 2.8   & 2.9   & \multicolumn{1}{c}{3.3 \color{OliveGreen}\footnotesize(+0.7)} \\
    SNB  & 3.9    & 7.1   & 9.3   & 84.1    & 6.9  & 59.9   & \multicolumn{1}{c}{28.5 \color{OliveGreen}\footnotesize(+2.3)} \\
    SOvR  & 51.3   & 60.9   & 5.8   & 54.3   & 14.9    & 49.6  & \multicolumn{1}{c}{39.5 \color{OliveGreen}(+29.9)} \\
    NCM  & 78.2  & 79.8  & 12.1 & 87.2  & 44.5   & 84.9   & \multicolumn{1}{c}{64.5 \color{OliveGreen}(+8.0)} \\
    CBCL  & 75.9  & 77.3 & 12.0  & 88.7   & 48.0  & 86.1   & \multicolumn{1}{c}{64.7 \color{OliveGreen}(+8.2)} \\
    SLDA & \textbf{89.9}  & \textbf{90.0}  & \textbf{50.8}  & \textbf{95.7}  & \textbf{90.8}  & \textbf{95.5} & \multicolumn{1}{c}{\textbf{85.5} \color{OliveGreen}(+8.8)} \\
    SQDA  & 85.5   & 84.0  & 48.7  & 88.8   & 67.1    & 82.7    & \multicolumn{1}{c}{76.1 \color{OliveGreen}(+5.1)} \\\hdashline
    iCaRL & 82.9   & 85.7  & 31.0  & 90.9   & 76.9   & 90.8   & \multicolumn{1}{c}{76.4 \color{OliveGreen}(+4.1)} \\
    \midrule
    \textbf{Avg} & 52.9  & 55.1   & 19.5  & 66.1   & 39.4   & 61.7  &  \\
  \bottomrule    
\end{tabular}%
}
  \label{tab:results_main_table_online_cl_methods}%
\end{table}%

\noindent\textbf{TAP vs.\ other pooling schemes} is evaluated in Tab.~\ref{tab:results_main_table_pooling_methods} for the best performing OCL method (SLDA).
STOCH and L2/L3 pooling underperform AVG.
RAP outperforms AVG and is comparable to MAX. 
AVGMAX is highly effective compared to AVG and MAX used alone. %
\begin{table}[tbp]
\setlength{\tabcolsep}{1.5pt}
  \centering
  \caption{Acc of SLDA under different pooling strategies on GSC (best in \textbf{bold}, runner-up \underline{underlined}).}
 \resizebox{\linewidth}{!}{%
    \begin{tabular}{lccccccc}
\toprule
    \multicolumn{1}{r}{} & W2V-B & W2V-L & Emf-B & HB-B & HB-L & HB-XL & \multicolumn{1}{c}{\textbf{Avg}} \\
    \midrule
    AVG \cite{lecun1998gradient}  & 82.4  & 81.6  & 23.2  & 94.2  & 85.5  & 93.3  & \multicolumn{1}{c}{76.7} \\
    MAX \cite{riesenhuber1999hierarchical} & 87.7  & 88.3  & 34.9  & 94.8  & 87.2  & 94.1  & \multicolumn{1}{c}{81.2} \\
    MIX (50\%) \cite{zhou2021mixed} & 87.8  & 87.3  & 31.1  & 94.7  & 87.3  & 94.0  & \multicolumn{1}{c}{80.4} \\
    STOCH \cite{zeiler2013stochastic} & 80.9  & 77.6  & 24.6  & 85.3  & 64.5  & 77.9  & \multicolumn{1}{c}{68.4} \\
    L2  \cite{feng2011geometric}  & 79.0  & 79.7  & 17.4  & 92.9  & 74.4  & 89.7  & \multicolumn{1}{c}{72.2} \\
    L3 \cite{feng2011geometric}   & 79.3  & 81.2  & 15.9  & 92.4  & 70.4  & 89.1  & \multicolumn{1}{c}{71.4} \\
    RAP (10\%) \cite{bera2020effect} & 86.5  & 86.9  & 36.3  & 94.8  & 87.5  & 93.8  & \multicolumn{1}{c}{81.0} \\
    AVGMAX \cite{monteiro2020performance} & \underline{89.1}  & \underline{89.6}  & 44.8  & \underline{95.2}  & \underline{89.1}  & \underline{94.7}  & \multicolumn{1}{c}{\underline{83.8}} \\
    iSQRT-COV \cite{li2018towards} & 80.3  & 80.3  & \textbf{55.1}  & 92.4  & 83.8  & 90.3  & \multicolumn{1}{c}{80.4} \\
    TSDP \cite{wang2021revisiting} & 83.9  & 83.6  & 32.4  & 94.4  & 84.9  & 93.9  & \multicolumn{1}{c}{78.9} \\
    TSTP \cite{wang2021revisiting} & 87.4  & 87.6  & 39.1  & 95.1  & 88.0  & 94.5  & \multicolumn{1}{c}{82.0} \\
    TAP (ours) & \textbf{90.0}  & \textbf{90.0}  & \underline{50.8}  & \textbf{95.7}  & \textbf{90.8}  & \textbf{95.5}  & \multicolumn{1}{c}{\textbf{85.5}} \\
    \midrule
    \textbf{Avg} & 84.5  & 84.5  & 33.8  & 93.5  & 82.8  & 91.7  &  \\\bottomrule   
\end{tabular}%
}
  \label{tab:results_main_table_pooling_methods}%
   \vspace{-0.1cm}
\end{table}%
To extract more useful temporal clues, covariance (iSQRT-COV) or variance (TSDP and TSTP) of features are used. iSQRT-COV is especially useful on Emf-B: overall, it outperforms AVG but not MAX. Using only standard deviation (TSDP) is less useful than using it in conjunction with AVG (\ie, TSTP). TSTP shows robust gains compared to AVG in every scenario by preservation of temporal statistics after pooling.
Our TAP moves from similar ideas and encodes further higher-order statistics in the pooled output. TAP is the best performing method on every architecture except for Emf-B where it ranks second. On average, it improves by relative 
\um{37.8\%}
compared to AVG, by 
\um{10.5\%}
compared to  AVGMAX and by 
\um{19.4\%}
compared to TSTP. We confirm the superiority of our TAP for EOCL KWS, which is achieved by extracting rich temporal dynamics from the single iteration over the input data.

\noindent\textbf{Is TAP adding large overhead?} 
Tab.~\ref{tab:ablation:order_momentums} studies Acc and relative \% increase of parameters ($\Delta_p$) compared to the pre-trained backbone when using TAP. We observe that: 1) ${R=5}$ always brings the highest Acc for all the 3 methods; and 2) $\Delta_p$ of TAP-SLDA (0.12\% for the highest Acc) shows only minimal increase compared to that of SLDA (0.10\%). \umb{On average, TAP increases training (inference) time by just 2.1\% (2.6\%).}

\begin{table}[tbp]
\setlength{\tabcolsep}{2.2pt}
  \centering
  \caption{Acc and increase of parameters (\%) over the backbone ($\Delta_p$). Metrics averaged over the $6$ networks on GSC. We use TAP with variable $R$ ($R{=}1$ is AVG). TAP has minimal footprint.}
  \resizebox{\linewidth}{!}{%
    \begin{tabular}{lcccccccccccc}
    \toprule
    \multirow{2}{*}{\textbf{TAP+}}& \multicolumn{2}{c}{$R=1$} & \multicolumn{2}{c}{$R=2$} & \multicolumn{2}{c}{$R=3$} & \multicolumn{2}{c}{$R=4$} & \multicolumn{2}{c}{$R=5$} & \multicolumn{2}{c}{$R=6$} \\
    \cmidrule(lr){2-3}\cmidrule(lr){4-5}\cmidrule(lr){6-7}\cmidrule(lr){8-9}\cmidrule(lr){10-11}\cmidrule(lr){12-13}
          & \multicolumn{1}{c}{Acc} & \multicolumn{1}{c}{$\Delta_p$} & \multicolumn{1}{c}{Acc} & \multicolumn{1}{c}{$\Delta_p$} & \multicolumn{1}{c}{Acc} & \multicolumn{1}{c}{$\Delta_p$} & \multicolumn{1}{c}{Acc} & \multicolumn{1}{c}{$\Delta_p$} & \multicolumn{1}{c}{Acc} & \multicolumn{1}{c}{$\Delta_p$} & \multicolumn{1}{c}{Acc} & \multicolumn{1}{c}{$\Delta_p$} \\\midrule
     \multicolumn{1}{c}{FT} & 1.6  & 0.01  & 1.5  & 0.01  & 1.5  & 0.02  & 1.5    & 0.02  & \textbf{3.8}  & 0.03  & 2.6   & 0.04 \\
    \multicolumn{1}{c}{NCM} & 52.1   & 0.01  & 62.8  & 0.01  & 63.5   & 0.02  & 64.0 & 0.02  & \textbf{64.5}   & 0.03  & 56.7   & 0.04 \\
    \multicolumn{1}{c}{SLDA} & 76.7  & 0.10  & 82.0  & 0.10  & 84.4  & 0.11  & 85.2 & 0.12  & \textbf{85.5}  & 0.12  & 84.9  & 0.13 \\\bottomrule
    \end{tabular}%
    }
  \label{tab:ablation:order_momentums}%
\end{table}%

\noindent\textbf{Randomly initialized weights.} 
Pre-training helps speech tasks but requires abundant data and computing power. Even in case of no pre-training or fine-tuning, using TAP alone still improves the accuracy thanks to the richer temporal statistics extracted from the scarce input data as we prove in Tab.~\ref{tab:ablation:pretraining}. 

\begin{table}[tbp]
\setlength{\tabcolsep}{1.7pt}
  \centering
  \caption{Acc w/o and w/ pre-training evaluated on GSC.}
  \resizebox{\linewidth}{!}{%
    \begin{tabular}{lcccccccccccc}
\toprule    
    \multicolumn{1}{r}{} & \multicolumn{2}{c}{FT} & \multicolumn{2}{c}{TAP-FT} & \multicolumn{2}{c}{NCM} & \multicolumn{2}{c}{TAP-NCM} & \multicolumn{2}{c}{SLDA} & \multicolumn{2}{c}{TAP-SLDA} \\
\cmidrule(lr){2-3}\cmidrule(lr){4-5}\cmidrule(lr){6-7}\cmidrule(lr){8-9}\cmidrule(lr){10-11}\cmidrule(lr){12-13}
\multicolumn{1}{r}{} & w/o   & w/    & w/o   & w/    & w/o   & w/    & w/o   & w/    & w/o   & w/    & w/o   & w/ \\
    \midrule
    W2V-B & 2.5   & 2.5   & 2.5   & 5.4   & 13.0  & 67.5  & 20.2  & 78.2  & 42.3  & 82.4  & 47.2  & 89.9 \\
    W2V-L & 2.5   & 2.5   & 2.5   & 6.1   & 14.0  & 69.5  & 21.7  & 79.8  & 43.2  & 81.6  & 48.1  & 90.0 \\
    \bottomrule
    \end{tabular}%
    }
  \label{tab:ablation:pretraining}%
\end{table}%

\begin{table}[!tbp]
\setlength{\tabcolsep}{2pt}
  \centering
  \caption{Acc in IID setup on GSC. Absolute improvement compared to class-IID setup is reported within brackets.}
  \resizebox{\linewidth}{!}{%
    \begin{tabular}{lcccccccc}
    \toprule
    \multicolumn{1}{r}{} & FT    & TAP-FT & NCM   & TAP-NCM & SLDA  & TAP-SLDA \\
    \cmidrule(lr){2-3}\cmidrule(lr){4-5}\cmidrule(lr){6-7}
    W2V-B & 44.6 \footnotesize(42.1) & 78.4 \footnotesize(73.0) & 67.5 \footnotesize(0) & 78.2 \footnotesize(0) & 82.5 \footnotesize(0.1) & 90.0 \footnotesize(0) \\
    W2V-L & 51.7 \footnotesize(49.2) & 81.1 \footnotesize(75.0) & 69.5 \footnotesize(0) & 79.8 \footnotesize(0) & 81.6 \footnotesize(0) & 90.0 \footnotesize(0) \\
    Emf-B & 13.9 \footnotesize(11.9) & 10.4 \footnotesize(8.9) & 8.8 \footnotesize(0) & 12.1 \footnotesize(0) & 23.2 \footnotesize(0) & 50.8 \footnotesize(0) \\
    HB-B & 86.2 \footnotesize(84.7) & 88.1 \footnotesize(86.6) & 83.9 \footnotesize(0) & 87.2 \footnotesize(0) & 94.2 \footnotesize(0) & 95.7 \footnotesize(0) \\
    HB-L & 52.6 \footnotesize(51.1) & 67.1 \footnotesize(65.6) & 46.7 \footnotesize(0) & 44.5 \footnotesize(0) & 85.5 \footnotesize(0) & 90.8 \footnotesize(0) \\
    HB-XL & 86.8 \footnotesize(85.3) & 88.1 \footnotesize(86.6) & 62.5 \footnotesize(0) & 84.9 \footnotesize(0) & 93.3 \footnotesize(0) & 95.5 \footnotesize(0) \\
    \hline
    \textbf{Avg} & 56.0  & 68.9 & 52.1  & 64.5  & 76.7  & 85.5 \\ \bottomrule
    \end{tabular}%
    }
  \label{tab:ablation:iid}%
   \vspace{-0.1cm}
\end{table}%

\noindent\textbf{Performance on IID case (no class ordering).}
Tab.~\ref{tab:ablation:iid} analyzes an IID setup where samples are seen at random (\ie, with no class ordering) by the CL model. On one hand, FT can greatly improve the Acc (as well as other models with high Forg on the class-IID setup of Tab.~\ref{tab:results_main_table_online_cl_methods_bis}) because the same class is presented to the learner at multiple stages during training, thus mitigating forgetting. On the other hand, we observe that methods relying on prototypes and covariance for inference are robust and not influenced by the non-stationarity of data in this setup, thus achieving results similar to the IID setup. Importantly, TAP is still effective even when no  consideration is taken: TAP-FT on IID data improves Acc by relative 
\um{29.3\%} 
over FT on average.

\noindent\textbf{Is larger feature space all we need?} We investigate whether increasing the size of the pooled output is enough to achieve the performance gains observed so far. We report results in Tab.~\ref{tab:ablation:feat_size} in terms of Acc and average feature size increment multiplier ($\Delta_{fs}$). AVG and MAX are the baselines with no increase in size (\ie, $\Delta_{fs}=1$). RAP extracts the top-k\% features: the large increase of feature size does not reflect into Acc gains. Motivated by the success of MAX and aiming at bringing temporal awareness into pooling, we considered a $2\!\cdot\!l$ window around MAX (\ie, MAXW$_l$). MAXW$_l$ can only match the results of MAX although showing higher feature size. On the extreme, FLAT flattens all features into a long 1D vector (\ie, no pooling). Despite drastic increase of feature size, these schemes cannot capture temporal dynamics of features. On the other hand, our method improves Acc thanks to the richer temporal statistics extracted from input features while maintaining practically unchanged the total number of parameters as shown in Tab.~\ref{tab:ablation:order_momentums}

\begin{table}[tbp]
\setlength{\tabcolsep}{1.5pt}
  \centering
  \caption{Acc of pooling methods which increase pooled feature size. Evaluation on the class-IID setup of GSC using SLDA as CL method. $\Delta_{fs}$: average feature size increment multiplier.}
  \resizebox{0.95\linewidth}{!}{%
    \begin{tabular}{lcccccccc}
    \toprule
    \multicolumn{1}{r}{} & W2V-B & W2V-L & Emf-B & HB-B & HB-L & HB-XL & \textbf{Avg} & $\Delta_{fs}$ \\\midrule
    AVG   & 82.4  & 81.6  & 23.2  & 94.2  & 85.5  & 93.3  & 76.7  & 1 \\
    MAX   & 87.7  & 88.3  & 34.9  & 94.8  & 87.2  & 94.1  & 81.2  & 1 \\\cdashlinelr{1-9}
    RAP 5\% & 85.7  & 85.8  & 35.1  & 94.5  & 86.8  & 93.8  & 80.3  & 26.8 \\
    RAP 10\% & 86.5  & 86.9  & 36.3  & 94.8  & 87.5  & 93.8  & 81.0  & 53.5 \\
    RAP 20\% & 85.7  & 85.9  & 36.3  & 94.5  & 86.8  & 93.8  & 80.5  & 107 \\
    MAXW\textsubscript{2} & 87.7  & 88.3  & 34.9  & 94.8  & 87.2  & 94.1  & 81.2  & 5 \\
    MAXW\textsubscript{5} & 85.8  & 86.5  & 34.9  & 94.7  & 87.1  & 93.9  & 80.5  & 11 \\
    MAXW\textsubscript{10} & 85.6  & 86.1  & 35.3  & 94.7  & 86.7  & 93.7  & 80.3  & 21 \\
    FLAT & 85.1  & 86.2  & 24.6  & 94.3  & 85.7  & 93.5  & 78.2  & 535 \\\hdashline
    TAP (R=2) & 87.4  & 87.6  & 39.1  & 95.1  & 88.0  & 94.5  & 82.0  & 2 \\
    TAP (R=3) & 89.3  & 89.2  & 47.3  & 95.5  & 89.8  & 95.4  & 84.4  & 3 \\
    TAP (R=4) & 90.0  & 90.0  & 49.7  & 95.6  & 90.4  & \textbf{95.5} & 85.2  & 4 \\
    TAP (R=5) & 90.0  & 90.0  & \textbf{50.8} & \textbf{95.7} & \textbf{90.8} & \textbf{95.5} & \textbf{85.5} & 5 \\
    TAP (R=6) & \textbf{90.1} & \textbf{90.2} & 47.8  & \textbf{95.7} & 89.9  & \textbf{95.5} & 84.9  & 6 \\\bottomrule
    \end{tabular}%
    }
  \label{tab:ablation:feat_size}%
\end{table}%

\begin{figure}[tb]
    \centering
    \includegraphics[trim=1.7cm 0.2cm 3cm 0.7cm, clip, width=0.85\linewidth]{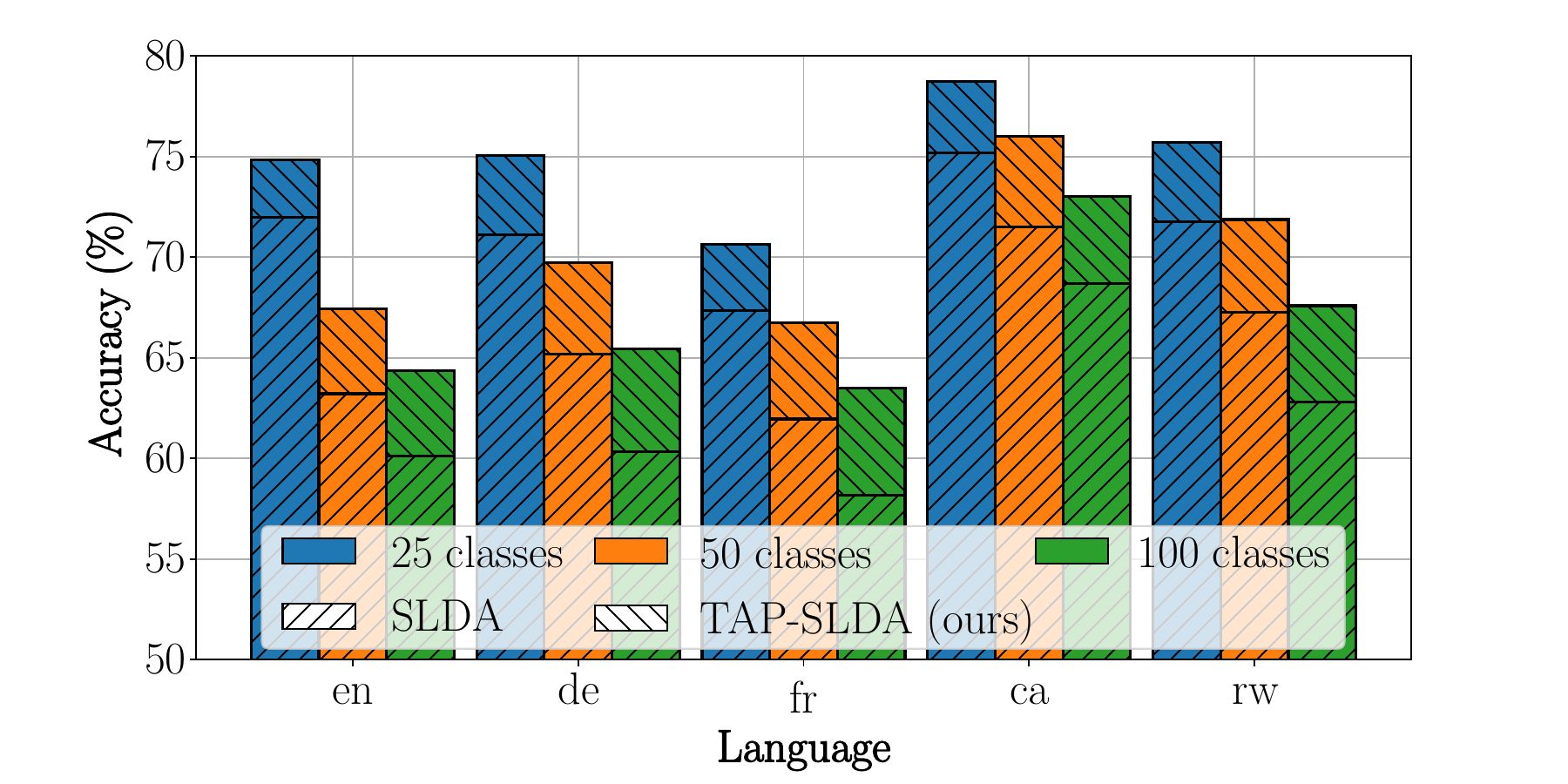}
    \caption{Acc of HB-B on MSWC micro-splits. Acc of FT averaged over class sets are respectively: 1.6, 1.5, 1.6, 1.4, 1.7.}
\label{fig:mswc}
 \vspace*{-0.2cm}
\end{figure}

\noindent\textbf{Personalization to other languages.} Fig.~\ref{fig:mswc} shows the best performing HB-B architecture (pre-trained on  unlabaled English data only) adapted online to recognize words from a different language. 
Acc is reduced compared to GSC, due to the harder MSWC benchmark; however, higher results can be achieved by multilingual pre-training \cite{wang2021unispeech}. Nonetheless, TAP-SLDA improves over the baseline on every language and class set (mean relative gain of 
\um{12.8\%}, and of 
\um{12.6\%} 
in the hardest case with 100 classes), therefore being extremely effective when domain of use changes, as it often happens for deployed KWS systems.

\section{Conclusion}
\label{sec:conclusion}
We proposed the first EOCL KWS system where only one sample is presented at a time on one single training epoch with a frozen backbone. We introduced a novel pooling mechanism (TAP) based on the first 5 moments of features to extract richer temporal-aware statistics. TAP provided the best results across different setups. %
\um{In particular, our TAP-SLDA 
achieves 95.7\% for EOCL KWS on GSSC, setting a new SOTA for the task.}

\clearpage
\bibliographystyle{IEEEtran}
\bibliography{strings_short,umbib}

\end{document}